# IMPACT OF MOBILITY ON THE PERFORMANCE OF MULTICAST ROUTING PROTOCOLS IN MANET

R. Manoharan and E. Ilavarasan

Department of Computer Science and Engineering
Pondicherry Engineering College, Puducherry
INDIA.

***ABSTRACT***

The advent of ubiquitous computing and the proliferation of portable computing devices have raised the importance of mobile ad-hoc network. A major challenge lies in adapting multicast communication into such environments where mobility and link failures are inevitable. The purpose of this paper is to study impact of mobility models in performance of multicast routing protocols in MANET. In this work, three widely used mobility models such as Random Way Point, Reference Point Group and Manhattan mobility models and three popular multicast routing protocols such as On-Demand Multicast Routing Protocol, Multicast Ad hoc On-demand Distance Vector Routing protocol and Adaptive Demand driven Multicast Routing protocol have been chosen and implemented in NS2. Several experiments have been carried out to study the relative strengths, weakness and applicability of multicast protocols to these mobility models.

***KEYWORDS***

*Mobile Ad hoc Network, multicast routing, mobility models, ODMRP, MAODV, ADMR.*

## 1. INTRODUCTION

Mobile ad hoc networks (MANETs) are self-organizing networks that do not require a fixed infrastructure. Two nodes communicate directly if they are in the transmission range of each other. Otherwise, they reach via a multi-hop route. Each MANET node must therefore be able to function as a router to forward data packets on behalf of other nodes [1]. Because of their unique benefits and versatilities, MANETs have a wide range of applications such as collaborative, distributed mobile computing (e.g., sensors, conferences), disaster relief (e.g., flood, earthquake), war front activities and communication between automobiles on highways. Most of these applications demand multicast or group communication.

Each of these applications can potentially involve in different scenarios with different mobility patterns, traffic rates dependent on the environment and the nature of the interactions among the participants. In order to thoroughly study the protocols for these applications, it is imperative to use the mobility models that accurately represent the mobile nodes which utilize the protocols. In this paper, it is proposed to analyze the performances of widely used multicast routing protocols namely Multicast Ad hoc On-demand Distance Vector (MAODV) routing protocol [2, 3], On-Demand Multicast Routing Protocol (ODMRP) [4, 5] and Adaptive Demand driven Multicast Routing protocol (ADMR) [6] against three different mobility model that characterize the realistic behaviours such as Random Waypoint, Reference Point Group and Manhattan mobility models.

Rest of the paper is organized as follows: Section 2 reviews the related work. Section 3 summarizes the Mobility Models that are considered in this paper. Section 4 explains the multicast protocols while Section 5 explains the experimental scenarios and methodology. Section 6 deals with experimental results. Finally, concluding remarks are given in section 7.

## 2. RELATED WORK

An extensive literature survey has been done to analyze the performance of routing protocols for various mobility models. Few researchers have carried out experiments to study the performance of unicast routing protocols such as DSR, DSDV, AODV and TORA in mobile environments [7]. Most of the initial research was using Random Waypoint as the underlying mobility model and CBR traffic consisting of randomly chosen source destination pairs. The protocols were mainly evaluated for packet delivery ratio and routing overhead. It was inferred that, the on-demand protocols such as DSR and AODV performed better than table driven ones such as DSDV at high mobility rates [7], while DSDV performed quite well at low mobility rates.

A comparison study of the two on-demand routing protocols namely DSR and AODV [2] was prepared with the packet delivery ratio and end to end delay metrics. It is inferred that DSR outperforms AODV in less demanding situations, while AODV outperforms DSR at heavy traffic load and high mobility. Another work proposed a framework to analyze the impact of mobility pattern on unicast routing performance of mobile ad hoc network [3], considering the Freeway mobility, Manhattan and RPGM mobility model.

The impacts of different mobility models on the performance of mobile IP multicast protocols are evaluated for two mobility metrics such as number of link changes and multicast agent density [8]. In [9], the authors have studied the effect of the different mobile node movement pattern in random-based mobility model group (Random Waypoint Mobility Model, Random Walk Mobility Model and Random Direction Mobility Model) on the performance of a unicast routing protocol AODV. The impact of different mobility models on mesh based Multicast Routing Protocols were analysed and presented in [10] by considering ODMRP and ADMR protocol under different mobility scenarios. A framework to analyse the impact of mobility model for unicast routing and on-demand routing is proposed in the literature [11, 12].

However, in the literature very few attempts were made to evaluate multicast routing protocols. The existing works do not capture the variety of mobility patterns likely to be exhibited by ad hoc applications and have not considered both tree based and mesh based multicast routing protocols for their study. Thus, in this work, we intend to study the performance of both tree and mesh based multicast routing protocol with three different mobility models.

## 3. MOBILITY MODELS

There are many mobility models proposed for use in MANET [13]. Out of the several mobility models [8], in this work, we consider three mobility models that are designed to capture a wide range of mobility patterns for ad-hoc applications. These models are briefly described in the following sections.

### 3.1. Random Waypoint Model

The Random Waypoint Mobility Model [8, 13] is a widely used mobility model, which imitate the natural entities move in extremely unpredictable direction and speed. In this model the Mobile Nodes (MN) includes pause times between changes in direction and/or speed. An MN begins by staying in one location for a certain period of time and then it move to another location by choosing a random destination and a speed that is uniformly distributed between minimum speed and maximum speed. Upon arrival, the MN pauses for a specified time period before starting the process again. In this model, the Mobile nodes are initially distributed randomly around the simulation area. This initial random distribution of Mobile nodes is not representative of the manner in which nodes distribute themselves when moving. The figure 1 shows the nodes moving in a simulation area with random speeds.

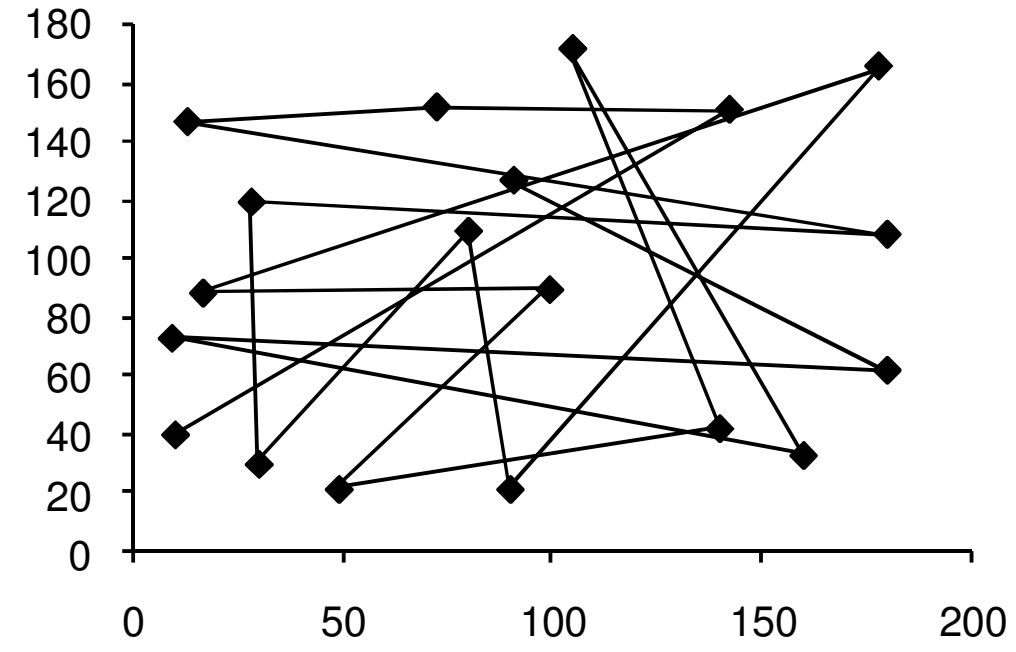

Figure 1. Travelling pattern of an MN using the Random Waypoint Mobility Model

### 3.2. Reference Point Group Model

Reference Point Group Mobility (RPGM) model [8] [13] [14], is a group mobility model which represents the random motion of a group of mobile nodes as well as the random motion of each individual node within the group [10]. Group movements are based upon the path travelled by a logical centre for the group. The logical centre for the group is used to calculate group motion via a group motion vector. The motion of the group centre completely characterizes the movement of its corresponding group of mobile nodes, including their direction and speed. Individual mobile nodes randomly move about their own pre-defined reference points, whose movements depend on the group movement. This mobility model is prevalent in many ad hoc applications which demand group communications.

### 3.3. Manhattan Model

The random way point and RPGM models are the random mobility models where the movement of mobile nodes are freely moving at any direction. In some mobile applications, the movement of mobile nodes follows the mobility pattern similar to the road maps. Thus Manhattan model [13] is also considered in this work. In the Manhattan model, the mobile nodes emulate the movement of nodes that are similar to the movement pattern on the streets defined by maps. In this model maps are used for the movement patterns. The map is composed of a number of horizontal and vertical streets. Each street has two lanes for each direction (North and South direction for vertical streets, East and West for horizontal streets). The mobile node is allowed to move along the grid of horizontal and vertical streets on the map. At an intersection of a horizontal and a vertical street, the mobile node can turn left, right or go straight. The figure 2 shows the map used for Manhattan mobility model.

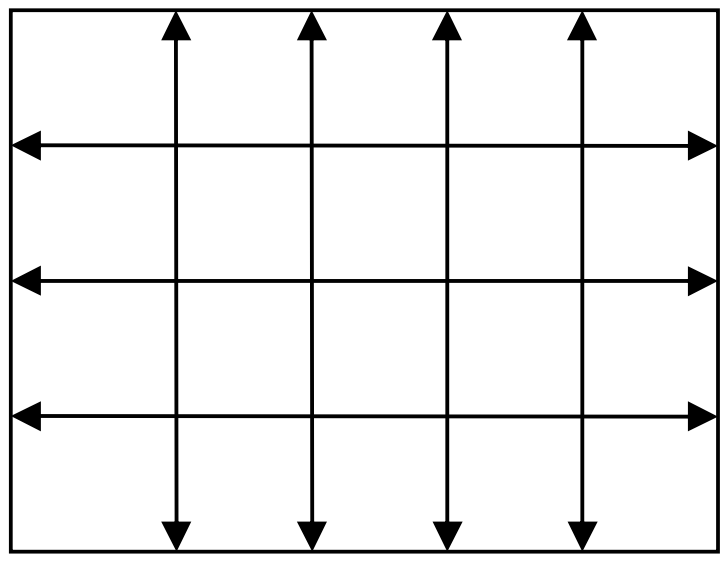

Figure 2. Map used in Manhattan Mobility Model

## 4. MULTICAST ROUTING PROTOCOLS

Multicasting is an effective way to provide group communication and it is very challenging in ad hoc networks due to the dynamic nature of the network topology. In this section, popularly used one tree based and two mesh based multicast routing protocols in mobile ad hoc network environment is described.

### 4.1. MAODV

MAODV protocol [2,3] is an extension of the AODV unicast protocol. This protocol discovers the multicast routes on demand using a broadcast route discovery mechanism employing the route request (RREQ) and route reply (RREP) messages. A mobile node originates an RREQ message when it wishes to join a multicast group, or has data to send to a multicast group but does not have a route to that group. Only a member of the desired multicast group may respond to a join RREQ. If the RREQ is not a join request, any node with a fresh enough route (based on group sequence number) to the multicast group may respond. If an intermediate node receives a join RREQ for a multicast group of which it is not a member, or it receives a RREQ and does not have a route to that group, it rebroadcasts the RREQ to its neighbors. As the RREQ is broadcast across the network, nodes set up pointers to establish the reverse route in their route tables. A node receiving an RREQ first updates its route table to record the sequence number and the next hop information for the source node. This reverse route entry may later be used to relay a response back to the source. For join RREQs, an additional entry is added to the multicast route table and is not activated unless the route is selected to be part of the multicast tree. If a node receives a join RREQ for a multicast group, it may reply if it is a member of the multicast group's tree and its recorded sequence number for the multicast group is at least as great as that contained in the RREQ. The responding node updates its route and multicast route tables by placing the requesting node's next hop information in the tables and then unicasts an RREP back to the source. As nodes along the path to the source receive the RREP, they add both a route table and a multicast route table entry for the node from which they received the RREP thereby creating the forward path. When a source node broadcasts an RREQ for a multicast group, it often receives more than one reply. The source node keeps the received route with the greatest sequence number and shortest hop count to the nearest member of the multicast tree for a specified period of time, and disregards other routes. At the end of this period, it enables the selected next hop in its multicast route table, and unicasts an activation message (MACT) to this selected next hop. The next hop, on receiving this message, enables the entry for the source node in its multicast routing table. If this node is a member of the multicast tree, it does not propagate the message any further. However, if this node is not a member of the multicast tree, it would have received one or more RREPs from its neighbors. It keeps the best next hop for its route to the multicast group, unicasts MACT to that next hop, and enables the corresponding entry in its multicast route table. This process continues until the node that originated the chosen RREP (member of tree) is reached. The first member of the multicast group becomes the leader for that group, which also becomes responsible for maintaining the multicast group sequence number and broadcasting this number to the multicast group. This update is done through a Group Hello message.

If a member terminates its membership with the group, the multicast tree requires pruning. Links in the tree are monitored to detect link breakages, and the node that is farther from the multicast group leader (downstream of the break) takes the responsibility to repair the broken link. If the tree cannot be reconnected, a new leader for the disconnected downstream node is chosen as follows. If the node that initiated the route rebuilding is a multicast group member, it becomes the new multicast group leader. On the other hand, if it was not a group member and has only one next hop for the tree, it prunes itself from the tree by sending its next hop a prune message. This continues until a group member is reached. Once separate partitions reconnect, a node eventually receives a Group Hello message for the multicast group that contains group leader information different from the information it already has. If this node is a member of the multicast group and if it is a member of the partition whose group leader has the lower IP address, it can initiate reconnection of the multicast tree.

## 4.2. ODMRP

A mesh-based demand-driven multicast protocol namely On-Demand Multicast Routing Protocol (ODMRP) [4, 5] which is, similar to Distance Vector Multicast Routing Protocol in wired network is considered. In this protocol, a source periodically builds a multicast tree for a group by flooding the control packet throughout the network. Nodes that are members of the group respond to the flood and join the tree. This is done by the source periodically flooding a JOIN QUERY message throughout the network. Each node receiving this message stores the previous hop from which it received the message. When a group member receives the JOIN QUERY, it responds by sending a JOIN REPLY to the source, following the previous hop stored at each node. Nodes that forward a JOIN REPLY create soft forwarding state for the group, which must be renewed by subsequent JOIN REPLY messages. If the node is already an established forwarding member for that group, then it suppresses any further JOIN REPLY forwarding in order to reduce channel overhead. The basic trade-off in ODMRP is between throughput and overhead. A source can increase throughput by sending more frequent JOIN QUERY messages. Each message rebuilds the multicast mesh, repairing any breaks that have occurred since the last query, thus increasing the chance for subsequent packets to be delivered correctly. However, because each query is flooded, increasing the query rate also increases the overhead of the protocol.

### 4.3. ADMR

The second protocol we consider is ADMR [6]. ADMR creates source specific multicast trees, using an on-demand mechanism that only creates a tree if there is at least one source and one receiver active for the group. Sources periodically send a network-wide flood, but only at a very low rate in order to recover from network partitions. In addition, forwarding nodes in the multicast tree may monitor the packet forwarding rate to determine when the tree has broken or the source has become silent. If a link has broken, a node can initiate a repair on its own, and if the source has stopped sending, then any forwarding state is silently removed. Receivers also monitor the packet reception rate and can re-join the multicast tree if intermediate nodes have been unable to reconnect the tree.

To join a multicast group, an ADMR receiver floods a MULTICAST SOLICITATION message throughout the network. When a source receives this message, it responds by sending a unicast KEEP-ALIVE message to that receiver, confirming that the receiver can join that source. The receiver responds to the KEEP-ALIVE by sending a RECEIVER JOIN along the reverse path. In addition to the receiver's join mechanism, a source periodically sends a network-wide flood of a RECEIVER DISCOVERY message. Receivers that get this message respond to it with a RECEIVER JOIN if they are not already connected to the multicast tree. Each node begins a repair process if it misses a defined threshold of consecutive packets. Receivers do a repair by broadcasting a new MULTICAST SOLICITATION message. Nodes on the multicast tree send a REPAIR NOTIFICATION message down its subtree to cancel the repair of downstream nodes. The most upstream node transmits a hop-limited flood of a RECONNECT message. Any forwarder receiving this message forwards the RECONNECT up the multicast tree to the source. The source in return responds to the RECONNECT by sending a RECONNECT REPLY as a unicast message that follows the path of the RECONNECT back to the repairing node. Nodes on the multicast tree also maintain their forwarding state. They expect to receive either PASSIVE ACKNOWLEDGEMENT (if a downstream node forwards the packet) or an EXPLICIT ACKNOWLEDGMENT if it is a last hop router in the tree. If defined thresholds of consecutive acknowledgments are missed then the forwarding node expire its state.

In all the above three protocols the overhead increases due to dynamic behavior of the node mobility resulting in link breakages.

## 5. IMPLEMENTATION

There are three techniques to evaluate the performance namely analytical modeling, simulation and measurement. In this work, simulation technique had been chosen because it is the most suitable technique to get more details that can be incorporated and less assumption is required compared to analytical modeling [15]. The performance evaluations of the protocols due to mobility have been carried out by implementing the protocols in NS2 simulator [16]. The implementation scenario is depicted in the Figure 3. The NS2 requires the mobility model and traffic pattern as an input. The mobility models have been generated using Java and the resultant file is converted into NS2 format. The traffic file is generated from the NS2 "cbrgen" tool [16]. The routing protocols are implemented using C++ and is set as parameter to NS2.

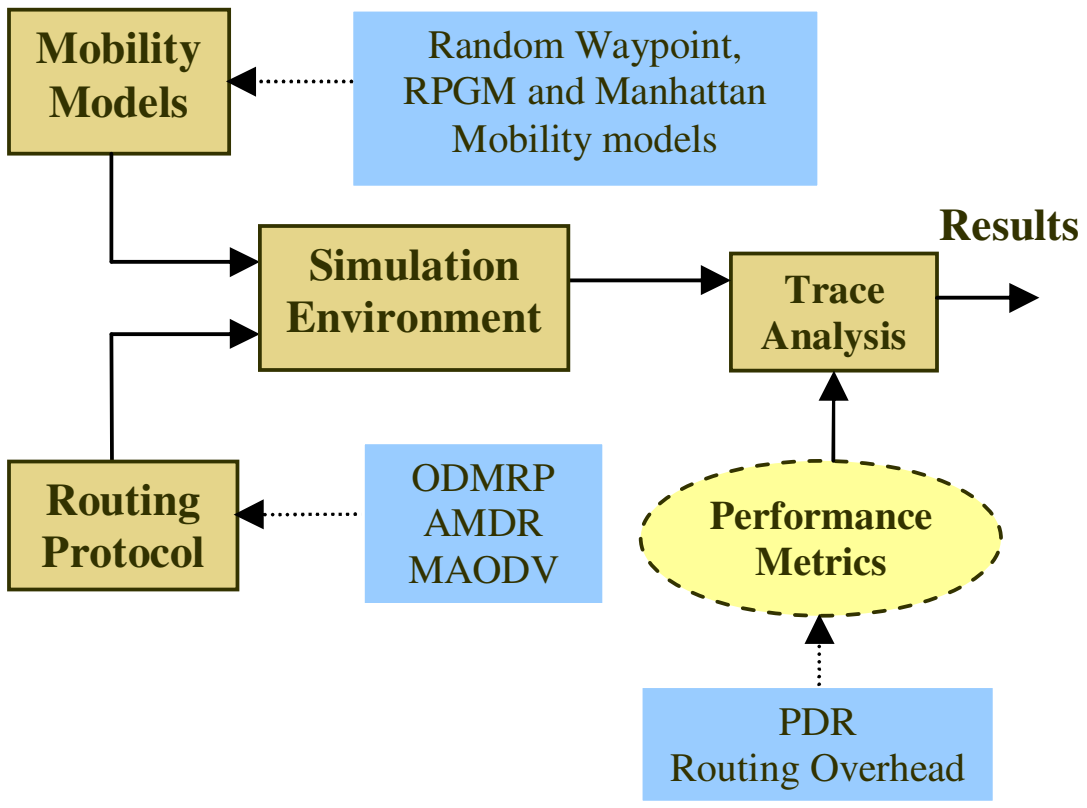

Figure 3. Implementation Design

The simulation outputs the trace files which are then analyzed using Perl. After extracting the various values from the trace file, the results were obtained. The results were averaged over several runs with the same simulation environment. The results and discussions are given in the subsequent section.

## 6. SIMULATION RESULTS AND PERFORMANCE COMPARISON

Our simulation models a dynamic mobile ad hoc network of 50 mobile nodes moved in an area of 1000m by 700m rectangular area. Each node has a uniform transmission range of 150m. The simulation has been run for each of the three mobility model with 10 multicast sessions and 10 nodes in each group. The multicast source and receiver nodes are selected at random. Multiple runs are conducted for different scenarios and the collected data is averaged over these runs. The mobility scenario generator produced the Random Waypoint, RPGM and Manhattan mobility patterns as required by the NS-2. Each run of the simulator accepts the scenario files that describe the exact motion of each node together with exact time at which each change in motion is to occur. We generated scenario files with varying node speeds. For all these scenarios MAODV, ODMRP and ADMR routing protocols were used for testing the performance variation due to mobility. The metrics used to measure the performance of routing protocols are

***Packet delivery ratio:*** The ratio of the number of packets originated by the application layer CBR sources to the number of packets successfully delivered to their CBR sink at the final destination.

***Normalized routing overhead***: It is the number of control packets transmitted per data packet received at the destination.

## 6.1. Packet Delivery Ratio

Out of the three routing protocols, it is observed that MAODV performs better than the other two protocols in term of packet delivery ratio which is shown in Figures 4, 5 and 6. Figures 7, 8 and 9 shows that the packet delivery ratios in Random Way Point model for the three protocols do not have sudden change when the speed of the mobile node increases. Thus Random Waypoint mobility model performs fairly well in all the three protocols.

MAODV has the highest packet delivery ratio when compared to ODMRP and ADMR. In MAODV there is significant decrease in the packet delivery ratio when the speed of the mobile node increases. It is obvious that when the mobile node moves with greater speed there are more chances for link breakage and result in less packet delivery ratio.

The throughput of ODMRP protocol depends purely on the mobility model and not much based on the speed of the mobile nodes. RPGM mobility model gives the better packet delivery ratio for ODMRP and the Manhattan model gives the worst packet delivery ratio because of the lower reachability. This ordering from the best to worst is roughly predicted by link changes.

ADMR is able to maintain high throughput for nearly all mobility models even as the speed increases. This is due to two mechanisms followed in the protocol. First, the forwarding nodes are able to initiate local repair mechanism of the multicast tree when the packet loss is occurring due to link breakage. Secondly, the receivers that are experiencing high packet loss can request the protocol to switch to flooding in order to control the packet loss.

### 6.2. Routing Overhead

From the figure 10 to figure 15, we observe that MAODV has the highest routing overhead when compared to ODMRP and ADMR among all the three mobility models. Generally, the routing overhead increases with the speed of the mobile nodes. RPGM model gives minimum overhead as it supports the group movement and hence ensures more reachability.

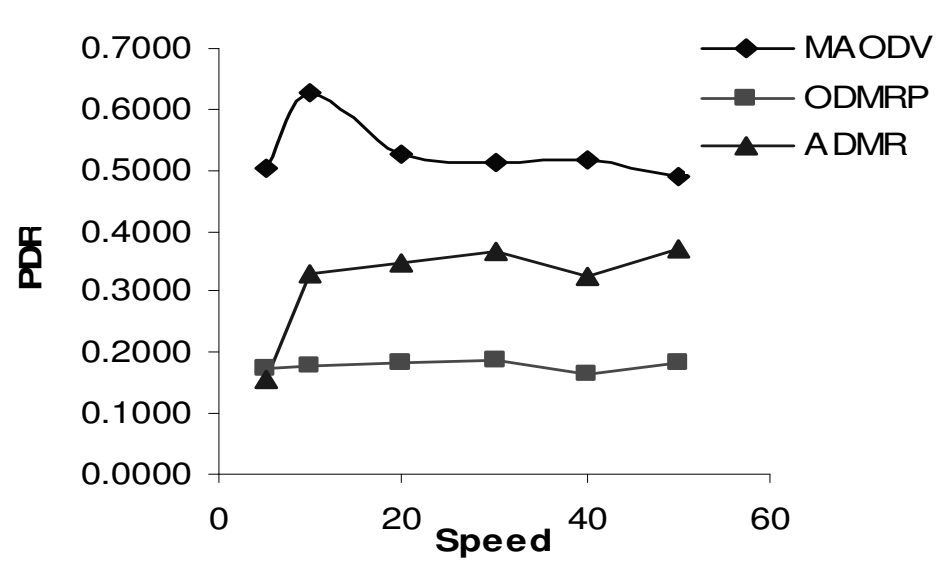

Figure 4. Manhattan Mobility Model

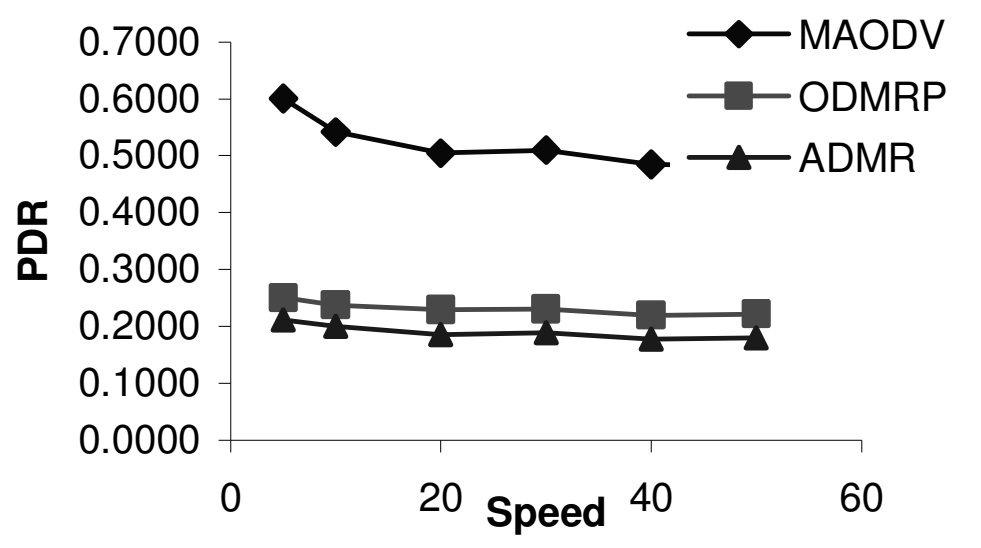

Figure 5. Random way Point Mobility Model

MAODV has the highest routing overhead for RPGM model than the other two models. This is because in group mobility model the link failures between different groups leads to changes with respect to the connection pattern. Hence the dynamic tree construction requires more control packets. ODMRP has the least routing overhead for RPGM model as it the mesh based routing protocol and it provides more than one path between two different nodes. ADMR has the minimum routing overhead compared to MADOV protocol over all the mobility models as it uses flooding at higher speeds.

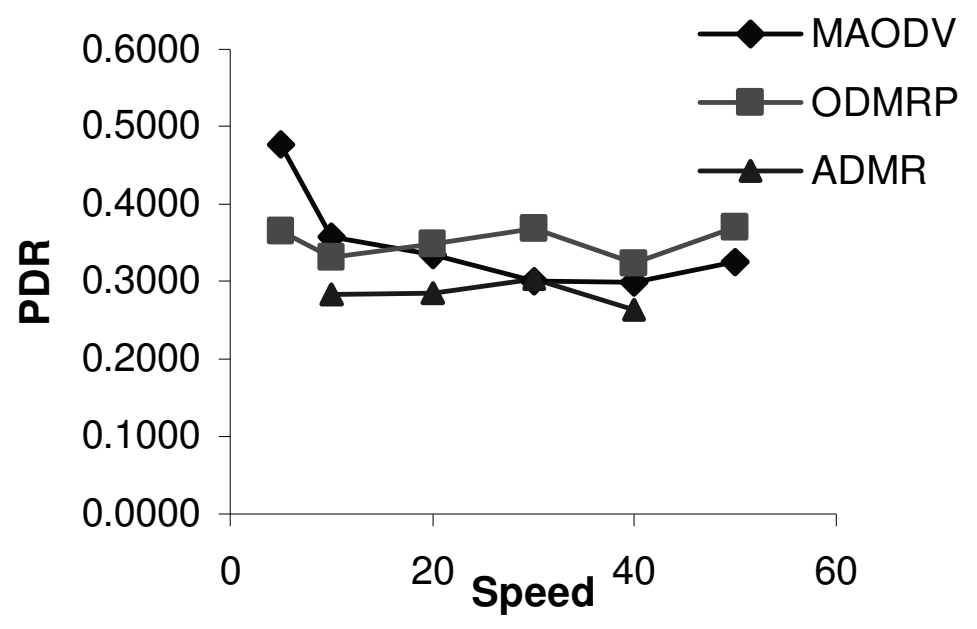

Figure 6. RPGM Mobility Model

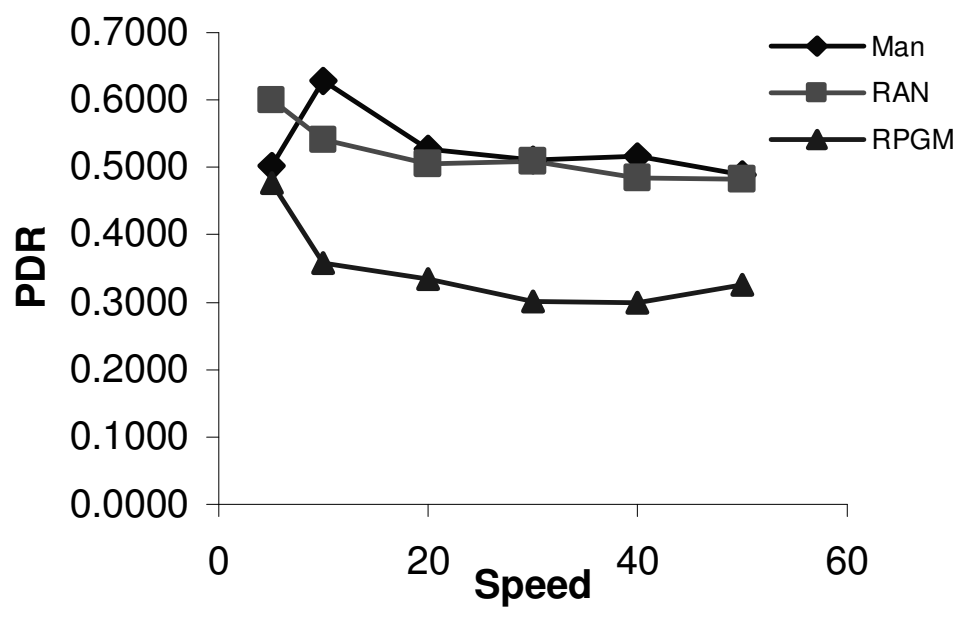

Figure 7. MAODV Protocol

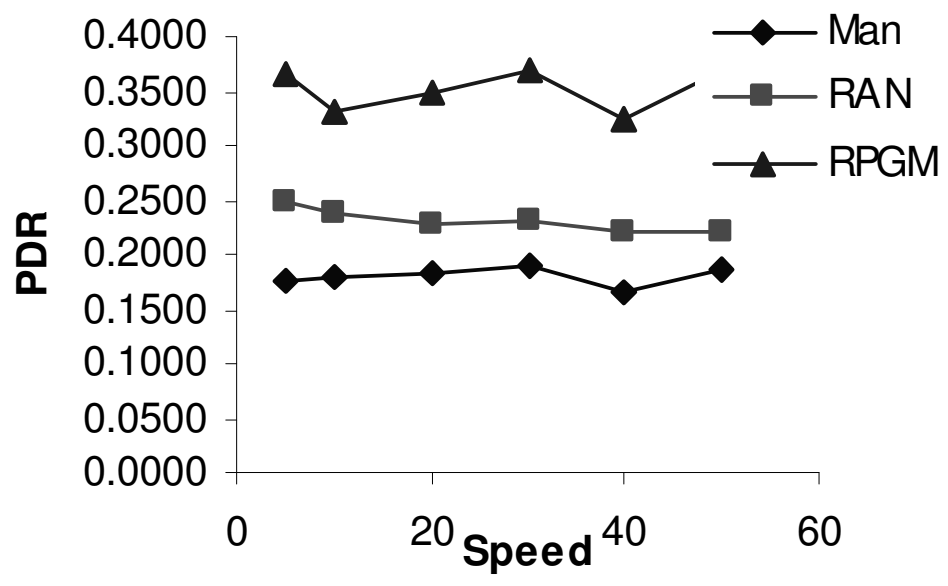

Figure 8. ODMRP Protocol

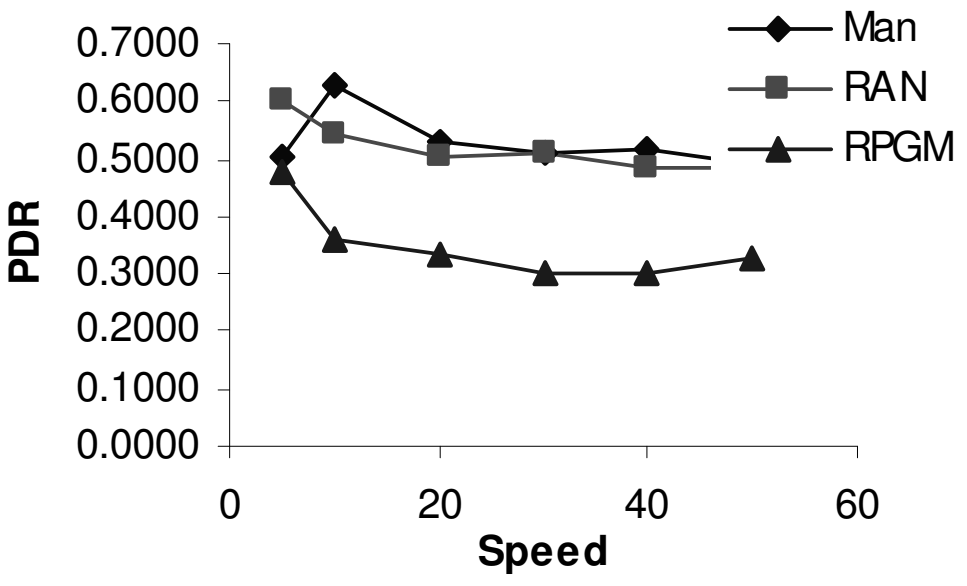

Figure 9. ADMR Protocol

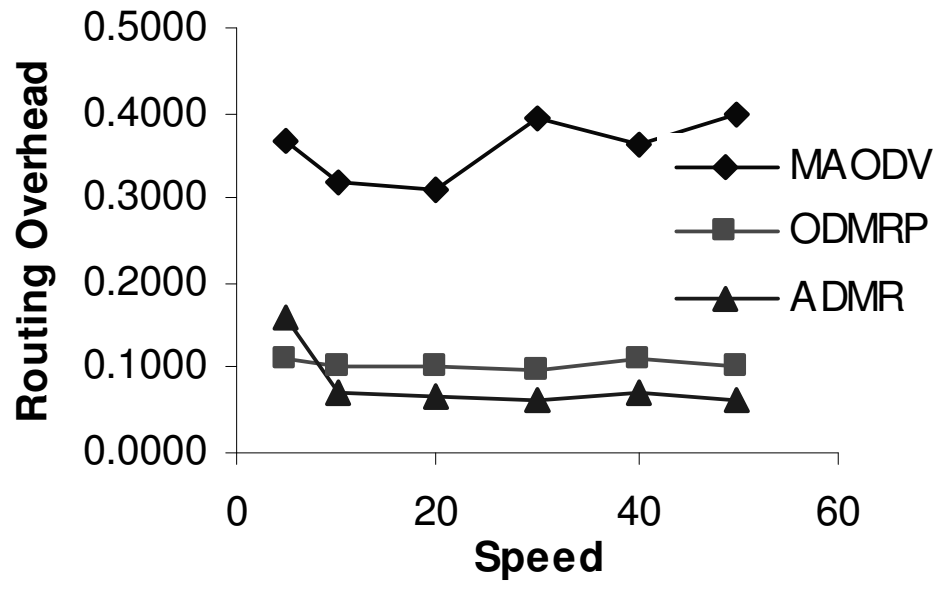

Figure 10. Manhattan Mobility Model

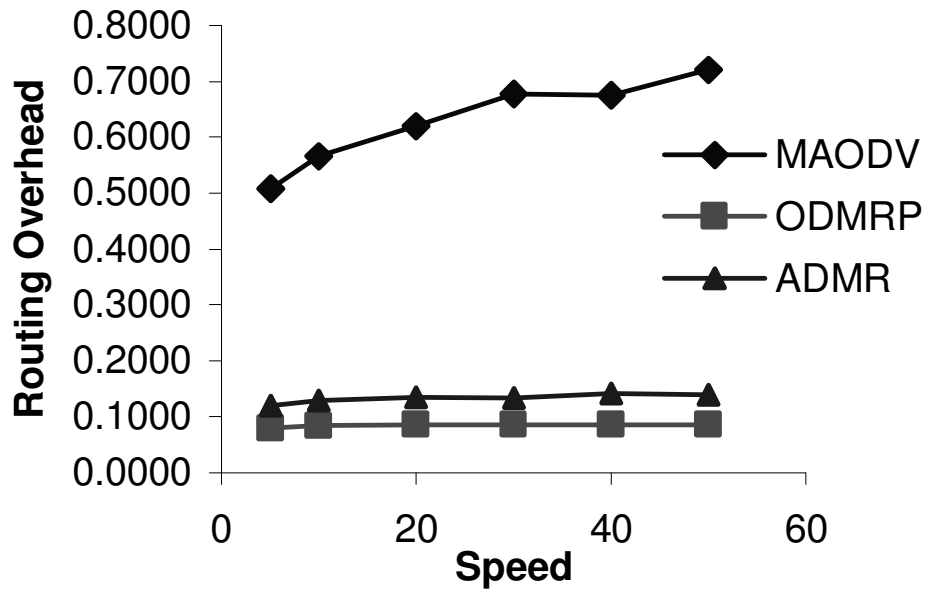

Figure 11. Random Way Point Mobility Model

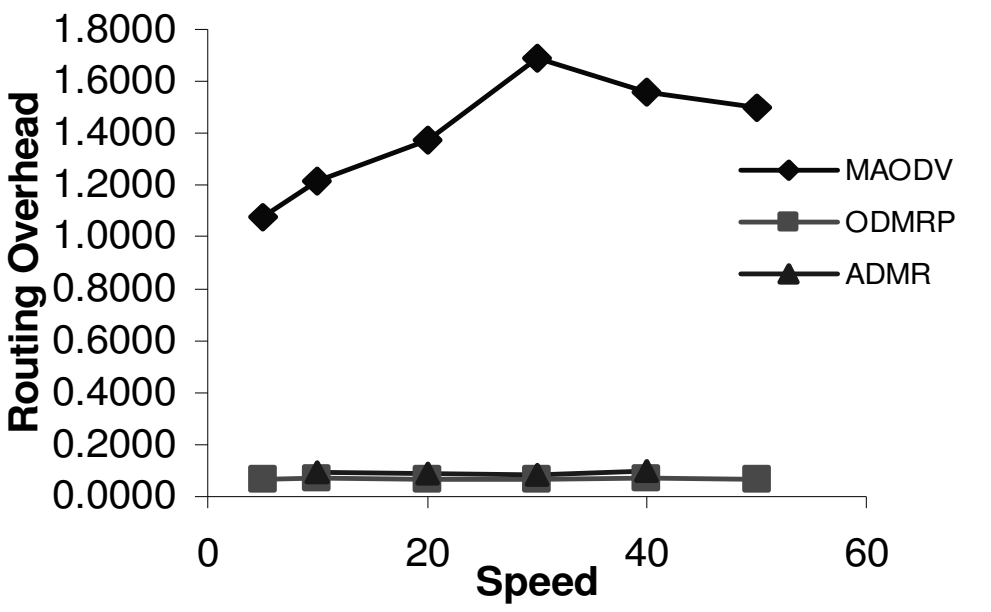

Figure 12. RPGM Mobility Model

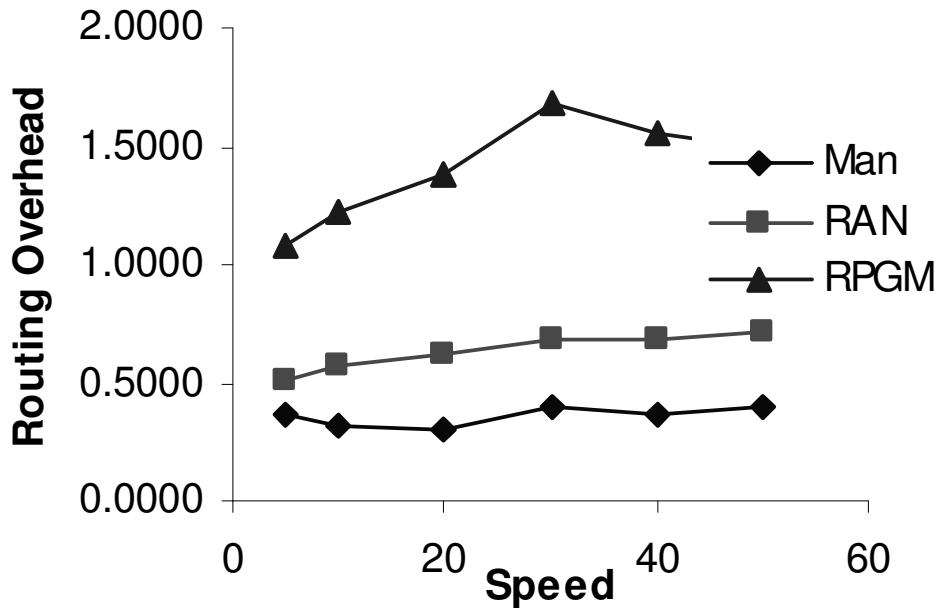

Figure 13. MAODV Protocol

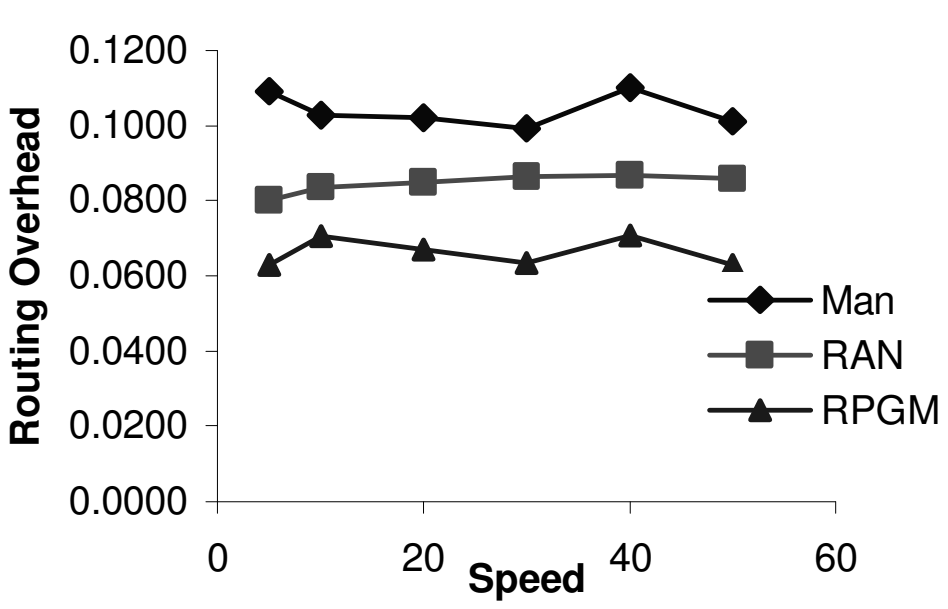

Figure 14. ODMRP Protocol

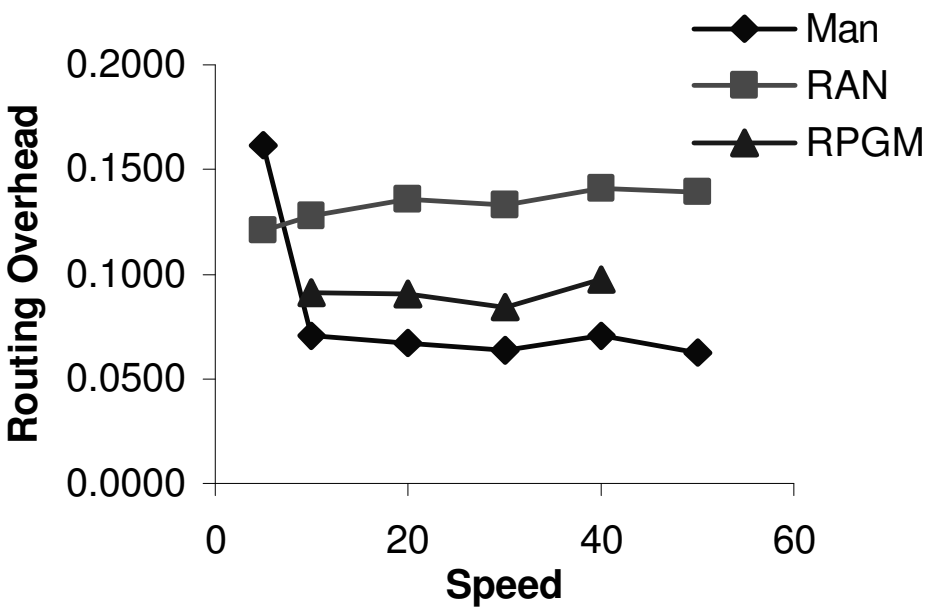

Figure 15. ADMR Protocol

## 7. CONCLUSION

In this paper, we analyzed the impact of mobility pattern on multicast routing performance of mobile ad hoc networks. We observe that in addition to the strengths and weaknesses of the individual multicast routing protocols, the mobility patterns does also have influence on the performance of the routing protocols. The connectivity of the mobile nodes, route setup and repair time are the major factors that affect protocol performance. This conclusion is consistent with the observation of the previous such studies on unicast routing protocols. There is no clear winner among the protocols in our case, since different mobility patterns seem to give different performance rankings of the protocols. This work can be further explored to study the impact of mobility on the performance of other multicast routing protocols. Several other parameters such as traffic patterns, node density and initial placement pattern of nodes may affect the routing performance and hence this work can be extended to investigate them further.

**Authors**

R. Manoharan is an Assistant Professor of Computer Science and Engineering at Pondicherry Engineering College, Puducherry, India. His area of specializations includes Mobile Networks and High Speed Networks. He has published more than twenty research papers in reputed conference and journals.

E. Ilavarasan is currently working as Assistant Professor in the Department of Computer Science and Engineering at Pondicherry Engineering College, Puducherry, India. He has published more than fifteen research papers in the International Journals and Conferences. His area of specialization includes Parallel and Distributed Systems, Computer Architecture and Design of Operating Systems.